Linking Things on the Web:
A Pragmatic Examination of Linked Data for Libraries, Museums and Archives.

Ed Summers
Library of Congress

Dorothea Salo
University of Wisconsin-Madison

License: [CC0](#)

**Abstract**

The Web publishing paradigm of Linked Data has been gaining traction in the cultural heritage sector: libraries, archives and museums. At first glance, the principles of Linked Data seem simple enough. However experienced Web developers, designers and architects who attempt to put these ideas into practice often find themselves having to digest and understand debates about Web architecture, the semantic web, artificial intelligence and the philosophical nature of identity. In this paper we will discuss some of the reasons why Linked Data is of interest to the cultural heritage community, what some of the pain points are for deploying it, and characterize some pragmatic ways for cultural heritage organizations to realize the goals of Linked Data with examples from the Web we have today.

*The Web is now philosophical engineering. Tim Berners-Lee. (Runciman, 2002)*

*The point of the Web arch is that it builds the illusion of a shared information space. Dan Connolly. (W3C, 2002)*

**Overview**

Since its introduction in 2006, Tim Berners-Lee's design notes on Linked Data have attracted increasing attention from the Web community (Berners-Lee, 2006a). Linked Data is a set of rules, or a pattern, for publishing data as hypermedia on the Web. It represents a simplification and distillation of the older, and conceptually more complex, Semantic Web project. Early adopters of Linked Data were typically associated with the W3C Semantic Web Education and Outreach (SWEO) Interest Group, that were interested in putting the ideas of the Semantic Web into practice, using openly licensed data. The publishing pattern soon saw interest from other sectors such as government, and the cultural heritage sector.

This paper examines some of the reasons why Linked Data principles have found a receptive audience in the cultural heritage community (libraries, archives and museums), as well as some of the challenges that publishing Linked Data poses for this community. Possible solutions to these problems are highlighted, based on the Web that we have, rather than a postulated future Web. In order to achieve these goals, a pragmatic and genealogical study of Web architecture is needed.

**Linked Data in a Nutshell**

Before examining why Linked Data is of interest to the cultural heritage sector it is first necessary to have a general understanding of what Linked Data is. Linked Data is best summarized by the "rules" for publishing Linked Data that were first introduced by Berners-Lee (2006a).

1. Use URIs as names for things
2. Use HTTP URIs so that people can look up those names.
3. When someone looks up a URI, provide useful information, using the standards (RDF*, SPARQL)
4. Include links to other URIs. so that they can discover more things.

These rules have been widely discussed, glossed and interpreted, for example by Heath and Bizer (2011). The rules' cornerstone is the assertion that HTTP Uniform Resource Identifiers (URI), or more commonly URLs, can be used to identify *documents* on the Web, as well as any

*thing* (real or fictional) that can be named in natural language, for example: the person Herman Melville, the novel Moby Dick, the place Nantucket Island, the Atlantic Ocean, etc. The notion that URLs are used to identify *resources* that are either *documents* or *real world objects* is central to the definition of Linked Data and also to the W3C's generalized description of Web Architecture [AWWW]. The distinction between types of resources is also central to some of the challenges related to the deployment of Linked Data, which will be discussed in more detail below.

Another key tenet of Linked Data is that when a URL for a *thing* is resolved by a Web browser, or any software agent using the Hypertext Transfer Protocol (HTTP), it receives data that has been modeled using the Resource Description Framework (RDF). Berners-Lee emphasizes HTTP URIs or URLs (which we will use hereafter) in his rules because URIs include identifiers that cannot be resolved easily, such as Uniform Resource Names, Handles, DOI, ISNI, etc. Being able to use the ubiquitous Internet, a global system of interconnected computer networks, to fetch data about a particular thing, is a key benefit of using URLs to identify things. However for the data to qualify as Linked Data in Berners-Lee's definition, a specific serialization of RDF data must be retrieved.

RDF is a Web oriented technology for conceptual modeling, that is similar to the more common entity relationship modeling and class diagramming techniques from the database world. At its core, RDF provides a grammar for making statements (or triples) about Web resources using URLs, that take the form of subject-predicate-object. Another subtle, yet important point is that RDF is a data model, not a data format in itself. However several data formats, or serializations, have been separately defined for RDF: RDF XML, RDFa, Turtle, N3, JSON-LD, n-triples, and n-quads. The variety of RDF serializations is why Berners-Lee used the notation RDF* in his third rule, to indicate that one or more of these RDF serialization formats must be used in Linked Data. The n-triples format is the easiest to understand since it lays bare both the subject-predicate-object grammar of RDF, as well as the central role that URLs play in the RDF data model. For example here is the RDF statement that a particular resource has the name Herman Melville:

```
<http://dbpedia.org/resource/Herman_Melville>
<http://xmlns.com/foaf/0.1/name> "Herman Melville" .
```

The requirement to use RDF was not always explicitly present in the initial publication of the rules in 2006, which simply said:

> *When someone looks up a URI, provide useful information. (Berners-Lee, 2006b)*

The text of the rules was edited in 2009 to be much more prescriptive about the use of RDF. The reasons for this small, yet significant change are rooted in Linked Data's relationship with the larger, and older Semantic Web project. The Semantic Web project began in 2001 (Berners-Lee, 2001) with similar overarching goals to Linked Data: to build a distributed, machine processable Web of data, with RDF is a foundational technology. Unlike Linked Data, the Semantic Web project aimed to roll out a stack of technologies to allow machine agents to discover new information using logical inference and rules. In this regard the Semantic Web inherited its goals from the much older research program of Artificial Intelligence (AI). Deployment of Semantic Web technologies to the Web community has proved slow, although there continue to be some small pockets of uptake in domains such as bio-informatics, where technologies like RDF and OWL have proved useful in an enterprise setting, if not on the open Web.

Berners-Lee's publication of Linked Data design issues distilled the essence of the Semantic Web project into a set of simple rules for Web publishers to use. The rules emphasise the value of building and navigating a Web of linked descriptions of things, rather than the logical inferencing technologies that can be used on an RDF dataset. The Semantic Web Education and Outreach group at the W3C helped popularize the term in order to solve the so called Semantic Web chicken-and-egg problem: it is difficult to motivate people to publish Semantic Web data if there are no applications that use it, and it is difficult to build useful Semantic Web applications if there isn't any data for them to operate on (Bizer, Heath, Ayers & Raimond, 2007). The tension between the W3C Semantic Web agenda firmly based on RDF and logical inferencing, and the more pluralistic principles of hypermedia on the Web is a topic we will return to later.

**The Cultural Heritage Sector**

Linked Data is a content agnostic technology, in that it can be used to publish all types of information: sociology, journalism, physics, philosophy, art, etc. However there has been significant interest in Linked Data from the cultural heritage community, as represented by libraries, archives and museums. The reasons for this interest are rooted in the institutional missions that these organizations share: to collect, preserve and provide access to our cultural heritage, and what they have in common with the goals of the Web, and specifically, Linked Data.

For as long as libraries, museums and archives have built collections they have also created descriptions of the cultural artifacts in their collections. Indeed, it is through their catalogs, inventories and finding aids that collections take shape, are managed and, ultimately, are used. The techniques, technologies and standards used to generate these descriptions have been wide ranging and often duplicative. Yet there has been a general trend towards collaborative sharing, both of the standards and technologies, as well as reusable elements of the descriptions themselves.

For example the Library of Congress began distributing catalog cards for books published in the United States in 1901. Card distribution lowered redundant cataloging work taking place in libraries around the country to describe the same book. Beginning in the late 1960s the Machine Readable Cataloging (MARC) standard was developed for representing cards in a card catalog, which allowed bibliographic information to be shared electronically, and cooperative cataloging computer networks such as OCLC saw their beginnings. The evolution of the Internet and later the World Wide Web pushed these descriptions online, initially to make distribution easier, and later so that information about collections could be obtained by researchers around the world.

Museums and archives have typically not shared complete descriptions as much as libraries, because they tend to collect unique objects instead of mass produced items like books. Archives specifically have a scale problem that libraries and museums do not: there is too much material to be described at the item level, so description tends to be done at the higher, more abstract level of series or subseries. Reuse and collaboration for archives and museums has centered on controlled vocabularies for identifying people, topics, places, and genre/form--shared descriptive practices. While the collections of libraries, archives and museums are quite different, many of these descriptive elements such as name authority files, thesauri and code lists have been shared and reused.

As these descriptions moved into a Web environment the term *metadata* was popularized, to reflect that descriptions were processable by a computer. The term is often used interchangeably with what is often more correctly called *data,* since all information is ultimately about something else, which is conceivable as information. In the mid to late 1990s the cultural heritage sector embraced XML technologies that were standardized by the World Wide Web Consortium. Efforts such as the Dublin Core Metadata Initiative and the Text Encoding Initiative provided collaborative spaces for the design and implementation of metadata practices. In a short period of time institutions and organizations created an astonishing number of descriptive standards, as illustrated by Riley and Becker (2009) in their visualization of the metadata universe.

As discussed by Lagoze (2006) and Tennant (2004) while standards such as XML made metadata widely processable using common tools (parsers, stylesheets) the diversity of document types and schemas has made interoperability difficult, which resulted in a multitude of mappings or cross-walking tools being created. Recently the object modeling approaches found in descriptive metadata practices such as Function Requirements for Bibliographic Records (FRBR), Preservation Metadata Implementation Strategies (PREMIS), the Dublin Core Abstract Model have paralleled the modeling efforts in the Linked Data community using RDF such as SKOS, Provenance, OAI-ORE and Open Annotation. In large part this movement towards RDF has been a recognition that defining document containers for metadata is insufficient for data modeling, and that entity relationship or object modeling is needed to see how these resources are identified and related together on and off the Web. Clarity about the identity and type of the resource being described is extremely important: for example, when describing the creator of a painting versus the creator of a JPEG representation of that painting. Unlike some metadata formats in use in libraries, attention to these details is an important component of RDF metadata, where the identify of the subject is central to the description. However, as we will discuss below, when these resources are placed on the Web, some ambiguity is re-introduced.

In addition to focusing on descriptive metadata, the cultural heritage community has developed standards for sharing or providing access to these machine readable descriptions. Beginning in the early 1980s work began on what would become Z39.50, an Internet based information retrieval protocol for remotely searching collections. In addition collection records were routinely shared as bulk datasets using the File Transfer Protcol (FTP) from services such as the Library of Congress Cataloging Distribution Service. As the Web become the dominant paradigm for publishing on the Internet in the late 1990s, Z39.50 was updated to use XML and HTTP as part of the Search/Retrieve by URL (SRU) standard. At the same time work on the Open Archives Initiative Protocol for Metadata Harvesting (OAI-PMH) provided a Web based synchronization mechanisms for digital repositories housed at libraries, archives and museums.

These repositories marked a departure for another reason, since they made not only descriptions of cultural artifacts available, but also the objects being described themselves (scholarly articles, digital surrogates for physical items, etc). At the same time, the OpenURL standard was developed for using URLs to connect licensed content found in publisher websites with the the abstract and indexing services used by researchers. In large part SRU, OAI-PMH and OpenURL use has been confined to the cultural heritage sector, as the larger Web saw similar developments such as search engine technologies (Sitemaps, OpenSearch), syndication technologies (RSS, Atom), and integration via RESTful Web Service APIs take hold. Cultural heritage organizations are now investigating Linked Data partly because it is perceived to be the latest in a series of alignments with Web architecture aimed at achieving better access to their collections. In addition the Linked Open Data project's focus on openly licensed content has caught the interest of cultural heritage organizations that want to support Open Access and Creative Commons, which encourage publishing with explicit licenses that allow others to know their right to reuse, repurpose that content.

The last and possibly most significant area of compatibility that libraries, archives and museums share with the Linked Data project is in the area of persistent identifiers, and digital preservation more generally. Cultural heritage organizations are entrusted with the safekeeping of cultural artifacts, for use now, and into the future. An essential part of this custodial responsibility is the proper identification of the artifacts that make up their collections. Historically this has included using identifiers to track physical items so that they can be located, for example: call numbers, barcodes, RFIDs, and box numbers. It also has included the unique identification of topics, people, places, languages, etc that help make up the description of an object. Authority control is particularly important in catalogs and indexes used by cultural heritage organizations for collocation to enable discovery, and use of a collection.

Prior to the Web, identification was achieved either through various shared code lists, or by formalized normalization rules for expressing a name, for example Mark Twain as:

> Twain, Mark, 1835-1910.

Since the arrival of the Web there has been an increasing awareness that opaque and resolvable identifiers for these entities are extremely useful, for example in the Virtual

International Authority File, which makes the following URL available for Mark Twain:

> http://viaf.org/viaf/50566653

To understand why the URL is preferable to the controlled string, consider the controlled name for the musician Robert Smith:

> Smith, Robert, 1959-

The "1959-" suffix is used to distinguish Robert Smith the musician from the many other people named Robert Smith in library catalogs. However when Robert Smith is no longer alive, his record will be updated to include the year that he died, similar to the form above for Mark Twain. At this point all the records that used the prior form would need to be updated, in order to keep the records properly collocated. The VIAF URL for Robert Smith does not share that problem, since it does not have to change when Robert Smith is deceased, and it has the added benefit of being resolvable by a Web browser:

> http://viaf.org/viaf/66490823/

At the same time, cultural heritage organizations have been historically wary of the URL, because of its perceived brittleness, or tendency to break. Indeed, significant research has been done in the digital library community on the phenomenon known as link-rot, which has estimated that 3% of URLs created in a given year will break (Nelson and Allen, 2002). Further studies by Sanderson, Phillips and Van de Sompel (2010) show how 28% of the papers referenced by a scholarly repository were lost. According to Ceglowski's (2011) study of URLs in the popular social bookmarking service Pinboard, links tend to die at a pretty steady rate: 25% of links are broken every 7 years. Distrust of the URL and the Domain Name System (DNS) upon which it is based has resulted in a series of identifier technologies being introduced such as the Digital Object Identifier (DOI), Handle, info URI, Archival Resource Key (ARK), Extensible Resource Identifier (XRI), and Uniform Resource Names (URN), Memento. These technologies have generally been advanced by the digital library and scholarly publishing community, who are typically looking at sustained use of identifiers over time. Yet these identifier technologies have not seen as wide adoption as the URL. Limited use has constrained the development of scalable software for using these alternate identifier technologies, which in turn has marginalized their community of users, and somewhat paradoxically potentially made them more prone to preservation failure. As a corollary to Linus' Law

> Given a large enough beta-tester and co-developer base, almost every problem will be characterized quickly and the fix obvious to someone. Or, less formally, "Given enough eyeballs, all bugs are shallow." (Raymond, 1999, p. 41) few eyes makes all bugs deep.

The ability of URLs to break in isolation, without breaking the Web completely, lowered the barrier to publishing on the Web, and allowed the Web to grow as an information space in ways that previous hypertext systems did not. In an effort to address the link rot issue, in 1998 Tim Berners-Lee published a W3C document describing a style of creating more durable URLs which he called "Cool URIs". The idea of Cool URIs (Berners-Lee, 1998) was aimed at improving the broken link problem without disrupting the ability of Web publishers to link to whatever they want on the Web, without needing to ask for permission. Cool URIs encouraged thinking of websites as namespaces in need of design, or as an information space that needs to be managed. Cool URIs also dovetailed nicely with Linked Data and the Semantic Web, which rely on stable Web identifiers as an essential building block.

So rather than there being a single reason, there are in fact a constellation of motivating factors for the cultural heritage sector's interest in Linked Data. It is because of their long standing interest in data modeling, identifiers, access technologies and preservation that libraries, archives and museums have found themselves looking at Linked Data as a possible technical solution to some of their problems, especially as these interests have come to be expressed in terms of the Web ecosystem (Baker et al., 2011). These are high hopes however, and we will discuss next where the aspirations of Linked Data meet the reality of deploying them in production settings.

**Publishing Linked Data**

As discussed above, a key aspect to publishing Linked Data is using URLs to identify *things* as well as *documents*. While it might sound simple, the distinction between things and documents is in fact quite subtle, and can prove problematic for cultural heritage organizations (and others) that wish to publish Linked Data. Before studying the difficulties of publishing Linked Data we will briefly examine why the Linked Data community sees the distinction to be so important, and what the recommended practice is for dealing with it.

In 2006, at roughly the same time that Tim Berners-Lee was authoring his Linked Data Design Notes, Leo Sauermann, Richard Cyganiak and Max Volkel began work on their *Cool URIs for the Semantic Web* (Sauerman & Cyganiak, 2006) which had the goal of providing simple guidance to Web publishers on how to publish data on the Semantic Web. The report was initially published as a technical memo by the Deutsche Forschungszentrum für Künstliche Intelligenz (German Research Center for Artificial Intelligence), and soon after was reviewed by the W3C Technical Architecture Group, and published as a W3C note. While it makes no mention of linked data, it became the seminal document for the Linked Open Data project of the Semantic Web Education and Outreach W3C group. The document was also important because it summarized literally years of discussion around perhaps one of the thorniest, and most contentious issues the W3C has faced during its existence: httpRange-14 (Berners-Lee, 2002a)

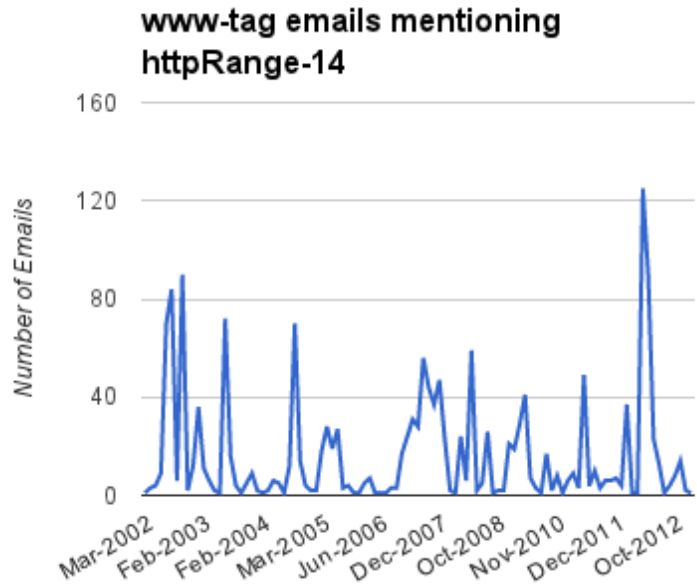

In March of 2002 the W3C Technical Architecture Group (TAG) created issue #14 in their issue tracking system, which had the title *What is the range of the HTTP de-reference function.* The issue has a long pre-history in the development of the Semantic Web technology stack, and was initially raised by Tim Berners-Lee on behalf of others in the Semantic Web community. The perceived problem was that the RDF based logical inferencing technologies being developed (RDFS, DAML+OIL, OWL) needed to distinguish between URLs that identified documents and URLs that identified real world entities. While this might sound complicated and esoteric it can be illustrated with a brief example, which assumes a basic understanding of RDF grammar of subject-predicate-object described above.

Since 1913 the New York Times has maintained an index to facilitate access to their articles, which was indispensable in the pre-Web era of microfilm based research. In 2009 a portion of this index was digitized and released as Linked Data on the Web (Sandhaus, 2009). For example, as a result of this effort the New York Times now publishes a URL that uniquely identifies Barack Obama:

```
http://data.nytimes.com/47452218948077706853
```

The goal of the Linked Open Data project is to build a globally distributed collection of reusable, openly licensed RDF data. When the New York Times initially released their data, this URL for Barack Obama returned RDF data that included an assertion that the data was licensed using the Creative Commons Attribution license:

```
<http://data.nytimes.com/47452218948077706853>
<http://creativecommons.org/ns#license>
```

<http://creativecommons.org/licenses/by/3.0/us/> .

Publishers of Linked Data are also encouraged to "link to other things" (Berners-Lee, 2006a), and specifically to link to things that are the same as the thing you are publishing, since links are the essential building block for building the Semantic Web and the Web in general. So the New York Times also asserted that their Barack Obama resource was the same as a resource published by the DBpedia project:

```
<http://data.nytimes.com/47452218948077706853>
<http://www.w3.org/2002/07/owl#sameAs>
<http://dbpedia.org/resource/Barack_Obama> .
```

The logical semantics of the sameAs predicate from Web Ontology Language (Dean and Schreiber, 2004) are such that when resources are said to to be the same, then anything that is true of one resource is necessarily true of the other. So for example if DBpedia asserts Barack Obama's birth date, an inference engine would automatically assign that birth date to the New York Times resource. This technique known as smushing is extremely useful when augmenting your data based on data elsewhere on the Web (Dodds & Davis, 2012). However any computer program that performs logical inference on the two New York Times and DBpedia assertions above would infer (based on the two RDF triples above) that the DBpedia resource was licensed with the Creative Commons license, when in fact DBpedia publish it using the Creative Commons Attribution Share Alike license. Additionally, the the DBpedia RDF for Barack Obama includes this statement

```
<http://dbpedia.org/resource/Barack_Obama>
<http://www.w3.org/1999/02/22-rdf-syntax-ns#type>
<http://xmlns.com/foaf/0.1/Person> .
```

So from the perspective of a machine agent the New York Times was also licensing the person Barack Obama. This particular situation has been described more fully by Halpin, Herman and Hayes (2009).

The Linked Data and Semantic Web communities' view on this particular inferencing problem is that it is the direct result of conflating the URL that identifies the *document* that describes Barack Obama with the URL that identifies Barack Obama himself. The proposed solution (and what the New York Times ended up implementing) was to use the document URL to assert the license, and to link to DBpedia using another URL.

```
<http://data.nytimes.com/47452218948077706853.rdf>
<http://creativecommons.org/ns#license>
<http://creativecommons.org/licenses/by/3.0/us/> .

<http://data.nytimes.com/47452218948077706853>
```

```
<http://www.w3.org/2002/07/owl#sameAs>
<http://dbpedia.org/resource/Barack_Obama> .
```

Cool URIs for the Semantic Web documents the various patterns sanctioned by the W3C resolution to the httpRange-14 issue, to help publishers share data on the Semantic Web while avoiding these sorts of logical quagmires. The patterns hinge on using distinct URLs to identify documents (or information resources) and real world things. The resolution to httpRange-14 specifically states that a URL for a real world thing must either end with a hash fragment (e.g. http://example.com/thing#) or the URL must return a HTTP See Other redirect to an information resource when it is resolved. While httpRange-14 was closed, consensus that it represents a viable solution to the issues it raised above is uncertain. Soon after the resolution in 2007, Issue 57 *Mechanisms for Obtaining Information About a Given URI* (Rees, 2007) was raised for addressing some concerns about the use of the 303 HTTP redirect, which later split into *Issue-62 Uniform Access to Server-provided Metadata* (Malhotra, 2009) and I*ssue-63 Metadata Architecture for the Web* (Masinter, 2009). As of this writing all these issues remain open.

**What is a Document?**

The publishing patterns in Cool URIs for the Semantic Web put a new emphasis on the thoughtful design of the URLs that web applications provide. However, the rules for publishing Linked Data require Web developers, designers and architects to do something they haven't really done before, namely to decide when a URL identifies an information resource or a real world thing. This seemingly simple task is fraught with practical and theoretical difficulties, especially for cultural heritage organizations.

For almost ten years, web application development has been greatly simplified through the introduction of web frameworks, or software libraries that provide the building blocks for a Web application. Among other things, these web frameworks have encouraged best practices in information architecture by providing mechanics for documenting and managing the URLs for resources that an application makes available on the Web. For example the popular Ruby on Rails web framework has the notion of routes that connect URL patterns to the controllers that generate a particular view or representation of a resource. Below is an example of a simple set of routes for a web application that makes information about books available:

```
ActionController::Routing::Routes.draw do |map|
    map.connect 'author/:id', 'author#view'
    map.connect 'publisher/:id', 'publisher#view'
    map.connect 'book/:id', 'book#view'
end
```

Without any knowledge of the Ruby programming language you can probably see that routes for three types of resources are defined: books, authors and publishers. You can also see that part of the route includes an `:id` which is passed to the relevant controller, in order to uniquely identify the entities in the application's database. If this Ruby on Rails application were deployed at http://example.com it would create URLs for individual books, authors and publishers that resembled:

- http://example.com/author/456
- http://example.com/publisher/789
- http://example.com/book/123

This sort of simple routing activity is not limited to Ruby on Rails; it is present in some form in hundreds of web application frameworks that are available for many programming language environments.

If a software developer, architect or designer wants to build a Linked Data application they must think about whether the URLs they are creating identify information resources or real world things. According to the Cool URIs for the Semantic Web document, the developer's answer to this question will define the form that the URL will take (end with a hash fragment), or the behaviour of the resource (HTTP 303 redirect) when it is accessed. In the case of Ruby on Rails this can be done by doubling the number of routes that are defined, a document and thing URL for each entity:

```
ActionController::Routing::Routes.draw do |map|
    map.connect '/doc/author/:id', 'author#doc'
    map.connect '/thing/author/:id', 'author#thing'

    map.connect '/doc/publisher/:id', 'publisher#doc'
    map.connect '/thing/publisher/:id', 'publisher#thing'

    map.connect '/doc/book/:id', 'book#doc'
    map.connect '/thing/book/:id', 'book#thing'
end
```

In addition to duplicating the routes, more code needs to be written in the controllers that define the logic for handling requests for the book as thing, which will redirect to the URL for the book as document. While certainly achievable, the point here is that the peculiarities of publishing Linked Data are typically not addressed in Web frameworks where the nature of the resource being an information resource or not isn't an affordance that is built into the technology. Unfortunately this isn't the only problem that Linked Data presents to Web publishers.

Consider for a moment whether the resources outlined above (authors, publishers and books) are information resources or not. The W3C define an information resource to be:

> By design a URI identifies one resource. We do not limit the scope of what might be a **resource**. The term "resource" is used in a general sense for whatever might be identified by a URI. It is conventional on the hypertext Web to describe Web pages, images, product catalogs, etc. as "resources". The distinguishing characteristic of these resources is that all of their essential characteristics can be conveyed in a message. We identify this set as "**information resources**." (Jacobs & Walsh, 2004).

Using this definition it's fairly clear that authors and publishers do seem to be real world things, but are books? Certainly it is possible to encode a book and send it as a message, say as a PDF document. So perhaps we do not need the URL route for the book as thing above? But then again a book is often an abstract Work that can have many translations, editions, and copies, so perhaps it is a thing afterall. This uncertainty is a theme that Wang (2007) explores in his study of the impact of the httpRange-14 decision. As a thought experiment Wang asks his reader to consider whether any of the following are information resources or not, and whether there is likely to be wide agreement between publishers on the Web:

- A book
- A clock
- The clock on the wall of my bedroom
- A gene
- The sequence of a gene
- A software
- A service
- A namespace
- An ontology
- A language
- A number
- A concept

Wang concludes that:

> *I doubt that anyone can give a definite answer. Hence, unless we can build an ontology that arbitrarily divides [all] conceivable things in the world into two groups and enforce people to use the classification, there is always the question - "what is an information resource?"*

Prior to the publication of the *Architecture of the World Wide Web* Tim Berners-Lee and others frequently referred to information resources as *documents*:

> *Formally, it is the publisher which defines the what an HTTP URI identifies, and so one should look to the publisher for a commitment as to the exact nature of the identity along these axes.*
>
> *I'm going to refer to this as a document, because it needs a term and that is the best I have to date, but the reader should be sure to realize that this does*

*not mean a conventional office document, it can be for example*

- *A poem*
- *An order for ball bearings*
- *A painting*
- *A Movie*
- *A review of a movie*
- *A sound clip*
- *A record of the temperature of the furnace*
- *An array a million integers, all zero*

*and so on, as limited only by our imagination. (Berners-Lee, 2002b)*

The terminology of documents situates Linked Data amidst an even older discourse concerning the nature of documents (Buckland, 1997), or documentation science more generally. Documentation science, or the study of documents is an entire field a field of study established by Otlet (1934), continued by Briet (1951), and more recently Levy (2001). As the use of computing technology spread in the 1960s documentation science was largely subsumed by the field of information science. In particular, Briet's contributions expanded the notion of what is commonly understood to be a document, by reorienting the discussion to be in terms of objects that function as organized physical evidence (e.g. an antelope in the zoo, as opposed to an antelope grazing on the African savanna). The evidentiary nature of documents is a theme that is particularly important in archival studies. Because of the nature of the material that they publish on the Web cultural heritage organizations (libraries, archives and museums) find themselves in the unfortunate position of constantly performing the information resource acid test. Professionals creating Linked Data web applications in these institutions not only have to bend the way that popular Web frameworks are used, but are routinely called upon to answer ontological questions about the nature of the resources they are publishing--whether they are an information resource or not.

**Genealogy of the URL**

In addition to the challenges of understanding the nature of information resources, Linked Data also highlights discontinuities or slippage in notions of Web architecture, and hints at mis-alignments between the people and organizations that determine the protocols that govern the Web. From the beginning, in Berners-Lee's (1989) proposal to build the World Wide Web at CERN, one can find the seeds for Linked Data. Berners-Lee's proposal includes a delightfully recursive diagram, which documents the relationships between itself (the proposal document), and other entities such as people (Tim Berners-Lee), organizations (CERN), topics (Hypertext), software projects (Hypercard), etc. So it is clear from this illustration that Berners-Lee foresaw, even as the Web was being born in 1990, that entities other than documents would be modeled on the Web.

And yet somewhere between 1990 and 1992 as the first standardization efforts around what would become known as the URL took shape, we see a shift towards defining document identifiers.

> *Many protocols and systems for document search and retrieval are currently in use, and many more protocols or refinements of existing protocols are to be expected in a field whose expansion is explosive.*
>
> *These systems are aiming to achieve global search and readership of **documents** across differing computing platforms, and despite a plethora of protocols and data formats. As protocols evolve, gateways can allow global access to remain possible. As data formats evolve, format conversion programs can preserve global access. There is one area, however, in which it is impractical to make conversions, and that is in the names used to identify documents. This is because names of documents are passed on in so many ways, from the backs of envelopes to hypertext **documents**, and may have a long life.* (Berners-Lee, Groff & Cailliau, 1992)

In 1994, after the idea for Universal Document Identifiers was socialized in the Internet Engineering Task Force (IETF), it became re-expressed as Universal Resource Identifier (URI), and Berners-Lee shifted to using the term object instead of document:

> *This document defines the syntax used by the World-Wide Web initiative to encode the names and addresses of **objects** on the Internet. The web is considered to include **objects** accessed using an extendable number of protocols, existing, invented for the web itself, or to be invented in the future. Access instructions for an individual **object** under a given protocol are encoded into forms of address string. Other protocols allow the use of **object** names of various forms. In order to abstract the idea of a **generic object**, the web needs the concepts of the **universal set of objects**, and of the universal set of names or addresses of objects. (Berners-Lee, 1994).*

Later that same year Uniform Resource Locators (RFC 1738) was authored by Tim Berners-Lee, Larry Masinter and Mitch McCahill. In this document the language switched from talking about objects to talking about resources. However RFC 1738 was largely quiet on the topic of what a resource was. In 1998 as Web technology was rapidly spreading, Tim Berners-Lee, Roy Fielding and Larry Masinter released Uniform Resource Locators (RFC 2396) which updated RFC 1738, and also provided a working definition for resource:

> *A resource can be anything that has identity. Familiar examples include an electronic document, an image, a service (e.g., "today's weather report for Los Angeles"), and a collection of other resources. Not all resources are network "retrievable"; e.g., human beings, corporations, and bound books in a library can also be considered resources.* (Berners-Lee, Fielding & Masinter, 1998).

RFC 2396 then stood for 7 years when it was obsoleted by the publication of URI Generic Syntax (RFC 3986) in 2005, by the same authors, which left the definition of resource largely the same:

> *This specification does not limit the scope of what might be a resource; rather, the term "resource" is used in a general sense for whatever might be identified by a URI. Familiar examples include an electronic document, an image, a source of information with a consistent purpose (e.g., "today's weather report for Los Angeles"), a service (e.g., an HTTP-to-SMS gateway), and a collection of other resources. A resource is not necessarily accessible via the Internet; e.g., human beings, corporations, and bound books in a library can also be resources. Likewise, abstract concepts can be resources, such as the operators and operands of a mathematical equation, the types of a relationship (e.g., "parent" or "employee"), or numeric values (e.g., zero, one, and infinity).* (Berners-Lee, Fielding & Masinter, 2005).

One can see echoes of the httpRange-14, which was being resolved at roughly the same time, reading in phrases such as:

> *A resource is not necessarily accessible via the Internet; e.g., human beings, corporations, and bound books in a library can also be resources.*

However the prescriptions found in the httpRange-14 resolution, to use URLs with hash fragments, or HTTP 303 redirects are absent from RFC 3986. And while *representations* are mentioned in RFC 2396, their role in Web architecture figures much more fully in RFC 3986, possibly as a result of Fielding's dissertation being published a few years earlier in 2000 (Fielding, 2000), which in turn influenced the W3Cs publication of *Architecture of the World Wide Web Volume 1* (AWWW) in 2004. It is interesting to note that while the AWWW introduces the term *information resource* it is not mentioned at all in RFC 3986, which is published a year later.

In this somewhat tangled genealogy of the URL it is tempting to read between the lines a bit, and observe a creative tension between two people, with competing pictures of Web architecture, who concentrate their work in different Web standards organizations. On one side we have Berners-Lee with the original notion of a Web of documents, that is to be augmented by a Web of things, a Semantic Web, which has been a project focus of the World Wide Web Consortium for over ten years. And on the other we have Fielding's notion of Web Architecture embodied in his dissertation on Representational State Transfer (REST), which was the result of working closely with Berners-Lee starting in 1993 on documenting and elucidating the Hypertext Transfer Protocol (HTTP), and the architectural constraints that it results from. Just as Berners-Lee's work was situated largely in the W3C, Fielding's work largely took place in the context of the IETF and the Apache Software Foundation which he used as a laboratory or training ground for his ideas. In truth one need look no further than Roy Fielding's response to the email from Tim Berners-Lee that raised the httpRange-14 issue in which he concludes:

> *RDF is broken if it cannot describe the Web in its entirety. The whole point in changing*

> *the model of the Web from a collection of Universal Document Identifiers that can be used to retrieve documents to one where Uniform Resource Identifiers are used to identify resources that can be accessed in the form of a representation through a uniform interface was so that we could accurately model how people have been actively using the Web since computational hypertext was introduced in 1993.* (Fielding, 2002)

or as Fielding previously wrote:

> *The early Web architecture defined URI as document identifiers. Authors were instructed to define identifiers in terms of a document's location on the network. Web protocols could then be used to retrieve that document. However, this definition proved to be unsatisfactory for a number of reasons. First, it suggests that the author is identifying the content transferred, which would imply that the identifier should change whenever the content changes. Second, there exist many addresses that corresponded to a service rather than a document—authors may be intending to direct readers to that service, rather than to any specific result from a prior access of that service. Finally, there exist addresses that do not correspond to a document at some periods of time, as when the document does not yet exist or when the address is being used solely for naming, rather than locating, information.* (Fielding & Taylor, 2002).

Somewhat paradoxically it was Fielding who sent the email on behalf of the W3C Technical Architecture Group that resolved the httpRange-14 issue, after which his own participation in the tag fell off noticeably.

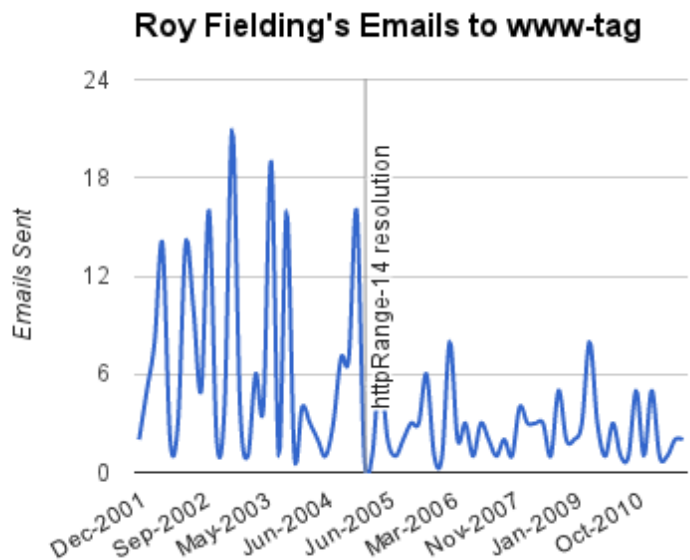

More recently, the Web Hypertext Application Technology Working Group (WHATWG) has been further standardizing the URL in the context of HTML 5 (Kesteren, 2012). In 2004 the WHATWG was formed, as a venue outside of the W3C, for browser implementers and others to continue work on HTML, since they observed that the W3C was more focused on development of XHTML and XML related technologies. The WHATWG's standardization work around the URL has been focused on simplifying the landscape by simply talking about URLs instead of URIs, IRIs, URNs; and providing a single algorithm and test suite for processing them. As with other work streams of the WHATWG, their work on the URL has been very much focused on improving Web application software. Interestingly, and perhaps wisely, Kesteren (2012) is completely silent on the nature of resources.

These technical specifications aside, It may very well be that the confusion around the nature of resources on the Web is best understood in the wider context of the philosophical nature of identity and language. Hayes and Halpin's (2008) work studying the dual role in location (retrieval) and naming (reference), provides an important backdrop to the technical debates about the use of URLs on the Web. The authors connect the Linked Data's agenda of naming things with URLs with the more general mechanics of naming in language:

> *When you use a language successfully we can usually assume people share agreement without relying on the impossible task of unambiguously connecting names to referents...Adding richer formal ontologies to a notion does not reduce ambiguity of reference, but it increases it by providing for finer ontological distinctions.* (Hayes & Halpin, 2008).

Seen in this light, it is conceivable that the introduction of a separate class of resources known as information resources has paradoxically made it more difficult to know what a resource is on the Web. Given Monnin and Halpin's (2012) continued work examining the philosophical shoulders that Linked Data stands on top of, makes the depth and difficulty of the debates around identity on the Web much more understandable.

**Conclusions and a Look Forward**

It could be argued that the issues above are merely stumbling blocks on the road to publishing Linked Data, and that there is no need to make mountains out of molehills. However, it is this author's position that these issues prove to be particularly thorny for the cultural heritage community, where what is meant by information resource takes center stage. That being said, Linked Data as a publishing practice offers very real benefits and opportunities for the cultural heritage community; especially as it embodies the architectural design constraints outlined by Fielding's Representational State Transfer. As an aside, it is significant that Fielding's pluralistic notion of Web architecture as an architectural style has been quietly, much more successful in guiding the design of Web applications and services, than prescriptive and often dogmatic rules about what format to publish data in, and how to name entities with URLs. The reasons for this discontinuity could form the body of another inquiry.

The first area that cultural heritage organizations can improve matters is in the area of RDF vocabularies. As Wang (2007) has noted, rather than expecting URLs and HTTP behaviors to differentiate between information resources and real world objects, the Linked Data community needs vocabularies that embody currently accepted notions of Web Architecture. Specifically, what is needed are RDF vocabularies that recognize that representations, not resources, are transmitted between the client and the server. The distinction between a resource and its representation is a fundamental part of Fielding's view of Web architecture, which has been reflected in the W3C's Architecture of the World Wide Web (Jacobs & Walsh, 2004.)

> *REST components perform actions on a resource by using a representation to capture the current or intended state of that resource and transferring that representation between components. A representation is a sequence of bytes, plus representation metadata to describe those bytes. Other commonly used but less precise names for a representation include document, file, and HTTP message entity, instance, or variant.* (Fielding, 2002, p. 126.)

If Web publishers need the ability to use RDF to describe the document-like properties of a representation then the semantics of these predicates need to make it clear when the representation is being described. For example, if it is important to be able to distinguish between the license for an HTML representation of a letter being published on the Web, and the license for the letter itself, then it would be useful to have a vocabulary term that let us make this more nuanced assertion, such as cc:representationLicense. Fortunately, as indicated by Scott (2010), the number of properties needed for describing the universe of representations is dwarfed by the number of properties needed for describing the universe of non-documents, so this need not be perceived to be a Sisyphean task.

Cultural heritage organizations stand to gain a great deal by engaging in the thoughtful management of URL namespaces, which is encouraged by Linked Data and information architecture more generally. Awareness of the types of resources that an organization is publishing on the Web, and where they can be reliably found in relation to each other is essential to their management over time (Archer, 2013). Terms of service documents which communicate the organization's intent when publishing resources are also important. For example, some institutions may need to allow contextual, exhibit-like or experimental portions of a website to change more fluidly, while resource specific views are more durable. Having a map of the existing URL namespaces can help cultural organizations plan for future work, and can also help insulate an organization from software or vendor dependencies and temporary branding decisions that can break the continuity of their online presence. The Web's continued growth is fueled by how easy it is to create and consequently break a link. Pools of well managed URL namespaces, that can be confidently linked to over time, are becoming the new libraries, archives and museums of the Web. Examples of these new libraries of the Web include large sites such as Wikipedia and the Internet Archive, as well as smaller sites like arXiv.org that moved from Los Alamos National Laboratory to Cornell University in 2001 without breaking their links (Steele, 2001). Relevancy ranking algorithms such as PageRank, which take into account the number of inbound links to a given resource, reward organizations that manage their URL namespaces, with traffic and attention. The longer a given Web resource lives, the more links will be created that point at it, and the higher its resulting PageRank relative to more evanescent resources. This attention to of cultural heritage materials, helps cultural heritage organizations fulfill their historic mission of making content available for use, and completes the virtuous circle of libraries, archives and museums on the Web.

Another opportunity area for cultural heritage organizations is the enhanced use of HTML. HTML's metadata facilities as well as the newer Microformats, Microdata, and RDFa specifications provide the mechanics for expressing metadata in Web pages themselves. This puts the ability to compose Linked Data in the hands of people who are comfortable adjusting HTML templates and configuring content management systems. It also simplifies deployment since it results in less Web resources needing to be managed, and also puts the data directly in front of users and automated agents that crawl the human readable Web. Using HTML as the container for cultural heritage data means that metadata practices are mainlined into Web publishing operations, where they can be used internally as well as externally for information management tasks such as federated search, link maintenance and metadata distribution. Vocabularies such as Facebook's Open Graph Protocol and Google's schema.org sidestep debates about the nature of resources and httpRange-14 by allowing Web publishers to assert what types of resources they are making available on the Web (e.g. books, scholarly articles, movies, etc), which is then used to provide additional contextual information in search results and social media environments. These user centered services that rely on structured metadata in HTML pages are likely to multiply, and are already incentivizing more organizations to publish structured metadata in their HTML. At the moment, the question of whether to use Microformats, Microdata or RDFa is largely a question of what tools and services you want to align with. The Web has thrived because it is a polyglot information space, and we can expect tools to continue to drive data format use.

> *The past is short and the future is long, and present determines how short and long they are. Thomas Aquinas.*

The Web is over twenty years old, and shows no sign of slowing down. There is increasing diversification in the standards bodies, and more importantly, the emergent qualities of praxis that help shape its topology. We can expect initial missteps in the area of httpRange-14 to be overtaken by simpler more intuitive ways of naming Web resources, which embrace their inherent ambiguity, rather than their logical entailments, and drive new functionality in search and social media technologies. The next twenty years will be interesting times for cultural heritage organizations as they adapt their historic missions to be expressed in terms of the Web: to serve as islands of persistence in larger ever expanding and changing sea of Web resources.